\documentstyle[preprint,aps]{revtex}

\begin{document}
\title{\bf Quantum Mechanics interpreted in
Quantum Real Numbers.}
\date{}
\author{John V Corbett\footnote{Department of Mathematics,Macquarie University,
N.S.W. 2109, Australia, Email address:jvc@ics.mq.edu.au}
and Thomas Durt\footnote{TENA, TONA Free University of Brussels, Pleinlaan
2, B-1050 Brussels, Belgium Email, address:thomdurt@vub.ac.be.} }
\date{}
\maketitle


\section*{abstract}
The concept of number is fundamental to the formulation of any
physical theory. We give a heuristic motivation for the reformulation
of Quantum Mechanics in terms of non-standard real numbers
called Quantum Real Numbers. The standard axioms of quantum mechanics
are re-interpreted. Our aim is to show that, when formulated in the
language of quantum real numbers, the laws of quantum mechanics
appear more natural, less counterintuitive than when they are presented
in terms of standard numbers.

PACS number(s): 03.65

\section*{Introduction}
In both classical and quantum physics,the states of a system are
represented by mathematical entities (points of the phase-space,
wave-functions) that ultimately consist of sets of real numbers.
These real numbers are either rational or arbitrarily well
approximated by rational numbers. The states are assumed to change
in time according to infinitely precise numerical laws, but
measurements only determine rational numerical values with finite
accuracy. Nonetheless, the accurate experimental confirmation of the
numerical predictions of quantum mechanics strongly encourages those
who believe in the basic role played by numbers in our universe and in
the potential for human beings to know and understand the laws that the
numbers obey. However we think that the often unstated assumption, "that
the elements of calculation are identical  with the elements of
observation" \cite{lindsay} is wrong. Our model \cite{adelman1} of 
quantum real numbers abandons this identification. Other abandonments are
well-known, for example,  Heisenberg's original paper on quantum mechanics
denied the assumption on the grounds that only relations "between
observable quantities" \cite{heisen} can be used. Our model does not
adhere to Heisenberg's requirement, on the contrary, in it
physical quantities take quantum  real numbers as their values even when
they are not observed. A more recent model that abandons this
identification is the non-commutative geometry \cite{connes} model of
A. Connes. Compared with it, our model is much less ambitious and requires
a less  radical change in the picture of the world because in it the
values of physical quantities are given by (commuting) Dedekind real
numbers, constructed as cuts in the rational numbers, even though not all
quantum real numbers exist to full extent.

As an example consider the position of a particle in a given state
at a given time. In the classical description, three real numbers
suffice to define the value of the position of a particle; in quantum
mechanics, the position of the particle is represented by a triplet
of self-adjoint operators. It usually is not acceptable to describe the
particle's position by a triplet of numerical values when the  particle's
state is represented by a wave-function. However it is generally conceded
that there is an average value for the position with a probability
distribution which is given by the modulus squared of the wavefunction in
position  space. Thus in the standard quantum mechanical picture a
quantum particle is not a material point but is associated with a cloud
of probabilities which is spread throughout space. Therefore quantum
physics seems to be non-local, an impression that has been confirmed, or
at least not contradicted, by all the experiments on Bell's inequalities.
Besides non-locality, which is revealed through the EPR paradox, the
measurement problem in quantum mechanics is at the source of several
paradoxical situations  (the Schr$\ddot{o}$dinger cat and the quantum Zeno
paradoxes for instance) that clearly illustrate the clash between
classical and quantum interpretations and ontologies.

In our view, understanding of the conceptual differences
between classical and quantum physics is improved by the recognition that
there can  be different realisations of real numbers determined by the
different theories\footnote {The logic of the non-standard quantum real
numbers is intuitionistic\cite{heyting}. For a
discussion of this aspect of the theory, and for instance, of the
reformulation of classical De Morgan's rules, see
\cite{adelman1}. The formulation of quantum real numbers in terms of
sheaves and toposes is given in \cite{adelman2}.}. To return to the
position of a particle, in our model its values are given by a triplet of
quantum real numbers, each of which relates, roughly speaking, to a
standard real number like a continuous function on an interval does to a
point in the interval. We shall give a fuller definition
of quantum real numbers later.  

We will not develop the quantum real numbers interpretation in a
strict axiomatic manner here.  We start by
accepting the standard Hilbert space mathematical structures that are used
in quantum theories but we do not accept their standard interpretation.

Schematically, our work is structured as follows: 
first the quantum real numbers are defined in basic postulate 0.
Then the prototype of a filtering or preparation procedure is taken to be 
the single slit experiment for the position of a particle.  This
experiment can be described classically when the quantum real number
associated to the position behaves classically in passing through the
slit. This is taken to mean that the square of this quantum number must
equal the quantum number associated to the square of the position to
within the order of a small positive standard real number $\epsilon$.
This situation is called the $\epsilon$ sharp collimation of the position.
If on passing through a slit there is strictly $\epsilon$ sharp
collimation for the particle's position then Theorem 4 shows that the von
Neumann transformation law holds for changes in the quantum real number
values of other quantities, up to the same $\epsilon$. A type of 
Heisenberg inequality for the widths of position and momentum slits is
obtained in Theorem 2.   

This analysis of the prototype motivates the introduction of 
the basic postulates of the quantum real number interpretation of
quantum mechanics.
 
Basic postulate 1 expresses the condition that the statistics of a
quantum experiment for any quantity is close to deterministic when
the experiment supports strictly $\epsilon$ sharp collimation for the
values of that quantity.

Next, we reinterpret, in terms of quantum real numbers, an argument due to
Goldstone \cite{goldstone,squires} and others to show that the
statistics relative to repeated measurements performed on identically
prepared systems always gives the conventional quantum probability
rules. That is, the Born probability rules\cite{vonneuneu2} hold.
The proof of Theorem 5 also assumes basic postulate 2 which
is equivalent to the quantum ergodic hypothesis. These two basic
postulates establish the  probabilistic features of
arbitrarily prepared quantum systems.

The basic postulates 3A and 3B express the persistence of measured
values, thereby restricting the range of application of these postulates.
The Luders-von Neumann transformation rule follows in Propositions 2 and
3.

The effect of this class of measurements of quantum real numbers is to
refine them to be a sharp quantum number which lie in the interval of
standard real numbers defined by  the resolution of the measuring
apparatus. Therefore in this interpretation the state of  a
system at a given time can be defined by the set of all the quantum real
numbers determined during the preparation/measurement process.
This corresponds closely to the classical concept except that quantum
real numbers are used instead of standard real numbers.

An example of how a combination of unitary, standard, dynamics and the
requirement of measured values to be unambiguously registered (revealed)
by a classical apparatus forces the measured values to be almost classical
quantum real numbers is proved in Proposition 4.

Next we consider single particle dynamics and the 19th century view that
matter is composed of atoms obeying Newtonian dynamics is modified in that
the values of the physical quantities are given by quantum real numbers
instead of standard real numbers. Basic postulate 4 asserts that the
dynamical equations of motion of a quantum system are given by
Hamilton/Newton's equations expressed in quantum real numbers.  Theorem 6
then  asserts that Heisenberg's operator equations of motion when
averaged over certain open subsets of state space closely approximate the
Newtonian equations for quantum real number defined on these open
sets. These open sets do not cover state space so that the approximate
equality cannot be extended to the whole of state space. 

Finally, we conclude by reformulating
some paradoxes in the language of quantum numbers.

\section{Quantum numbers-the one slit experiment.}
\subsection{Quantum numbers-the basic postulate 0.}
Consider a measurement of position. The passage
from one to three dimensions does not bring any new insight into the
problem, so that we will consider only the measurement of one
coordinate of position, let us say, the projection of the position along
the Z-axis. Our treatment is not relativistic so that
time will always be treated as an external parameter in the following.
The position of a classical particle (material point) along the Z-axis is
expressed by one standard real number: $z$.

Let $\Sigma$ be the set of
all density matrices
$\ \hat \rho$, where $\ \hat \rho$ is a positive-definite bounded
self-adjoint operator on ${\cal L}_{\rm
2}(\bf R)$ of trace 1. In conventional quantum mechanics, every
particular preparation procedure corresponds to a particular choice
or determination of a $\hat \rho$ or of a subset of them. A priori,
before we prepare the particle, we may be as ignorant of the states
$\hat \rho$ as we are of the values of
the position $z$ of the particle. Nevertheless in our model we 
assume that the particle has a set of states associated with it at
all times, even when we don't know what they are. Furthermore all
measurements of the position $z$ yield standard real numbers. We take
this to mean that while the position is associated with a self-adjoint
operator $\hat Z$ on  ${\cal L}_{\rm 2}(\bf R)$, its values are given by
quantum real numbers of the form $z_Q(U)$ with:

\begin{equation} z_Q(U)\ = \ \{Tr(\hat{\rho} \hat{Z});
\ \hat{\rho}\in{U}\}, \end{equation}
where $U$ is an open subset of $\Sigma$.  The open subsets of $\Sigma$
are defined by the weakest topology that makes 
$Tr(\hat \rho.\hat M)$ continuous as a function from $\Sigma$ to the
standard real numbers $\bf R$ for any linear operator $\hat M$ that is
self-adjoint and continuous. Here continuous means either bounded on
${\cal L}_{\rm 2}(\bf R)$ or continuous in the standard countably normed
topology on the Schwartz space ${\cal S}(\bf R)$. This topology is studied
in\cite{adelman2}; we shall call it the standard topology on $\Sigma$. 
Quantum real numbers of the form $z_Q(U)$ are real
numbers in the sense of Dedekind \footnote{Dedekind real numbers are
obtained as cuts in the  set of rational numbers $\bf Q$ . A cut is a
division of $\bf Q$ into two  classes $\bf L$ and $\bf R$, with every
rational in $\bf L$ less than every  rational in $\bf R$ . When $\bf L$
does not have a largest member and $\bf R$ does not have a smallest
member the cut defines an irrational number. For the quantum real number
$M_{Q}(U)$, where $U$ is an open subset of $\Sigma$ and $\hat M$ is a
self adjoint operator, the cut is defined by sections over subsets $W_j$
of an open cover of $U$ with the rationals $\bf Q_{Q}(W_j)$ given by
locally constants functions over $W$. Locally constant functions are
constant globally only when defined on a connected set
\cite{maclane,stout}.}
\cite{adelman2}.

We assume that preparation processes determine open sets $U$ and not
single states  $\hat \rho$. 

Thus we impose
the following definition of quantum number in the form of a
postulate:

{\bf Basic postulate 0:}

The values of a physical quantity are represented by quantum real
numbers of the form
$M_Q(U)\ =\ Tr(\hat \rho.\hat M)_{\hat
\rho\
\in\
U}$, where $U$ is an open subset of
the set of density matrices $\Sigma $,
and $\hat M$ is a self-adjoint, continuous linear operator. Furthermore,
every physical quantity has a quantum real number value at all
times.

Next we consider what happens when we
measure the quantum number $Z_{Q}(U)$ by letting the particle pass
through a slit. According to de Broglie every measurement is in the last
resort a measurement of position so that this will be a paradigm for the
measurement process.

\subsection{\bf The one slit experiment and sharp collimation.}

Assume that we can produce particles and prepare them to
pass through a rectangular slit in a vertical barrier. We assume that
the geometry of the experiment is such that the slit is infinitely
extended horizontally.  Let us denote by $z_1$ and
$z_2$ the Z-coordinates of the lower and upper extremities of the slit.

In classical mechanics, particles are assumed to behave as material
points and when the preparation is sufficiently accurate it is possible
to assign a unique well-defined trajectory to the particle. From
the knowledge of its initial position and velocity, one can in principle
deduce whether or not the particle will pass through the slit. Positions
are thus sharp numbers at each time, where sharpness is a measure of
the accuracy of our control and knowledge of
the experimental conditions. Obviously, for a classical particle, at the
time of the passage through the slit, its
position $z$ along the Z-axis lies between
$z_1$ and $z_2$ .

In the analogous quantum situation, we describe the passage
through a slit as a preparation process in terms of quantum
numbers. In the following treatment, all values are given at
the time of passage through the slit. If
$W$ is any open subset of $\Sigma $.  A $W$-prepared particle passes
through the slit if for all $\hat \rho$ in $W$, $Z_{Q}(\hat \rho )\ =\
Tr(\hat \rho.\hat Z)\ \in \  ]z_1,\ z_2[ $.
That is, in terms of the quantum number $Z_Q(W)$, if
 $Z_Q(W)\ \in \  ]z_1,\ z_2[$.

Furthermore, in classical mechanics, we are free to measure arbitrary
functions of the coordinates of the particles. For instance, when
$z_1$ and $z_2$ are positive numbers, instead of the coordinate
$z$ of the particle inside the slit, one could measure its square $z^2$.
In terms of quantum numbers, the situation is ambiguous
because, in general, for each $\hat \rho$, $ (Z_Q)^2(\hat \rho)\
=\ Tr^2(\hat \rho.\hat Z)$ differs from $(Z^2)_{Q}(\hat \rho) \
=\ Tr(\hat \rho.\hat Z^2)$, even when $ Z_{Q}(\hat \rho)\
=\ Tr(\hat \rho.\hat Z)\ \in \  ]z_1,\ z_2[$. We can use the difference
between these numbers to measure the departure from classical
behaviour, with the parameter $s(Z)$ defined, for each $\hat \rho$, 
as follows:

\begin{equation} s(Z)\ (\hat
\rho)\ =\ \sqrt{Tr(\hat
\rho.\hat Z^2)\ -\   Tr^2(\hat
\rho.\hat Z)}.\end{equation}

The Dedekind number $s(Z)(U)$ is defined pointwise on $U$. For any
open set $U$, $s(Z)(U)$ has the dimension of a length. We claim that
when $s(Z)(U)$ is much smaller than the width of the slit, $Z_{Q}(U)$
behaves as a standard real number.  The dimensionless ratio between
$s(Z)(U)$ and the width of the slit $(z_2\ -\ z_1)$ provides a 
measure of the departure from classicality that characterizes the passage
of a quantum particle through the slit. We introduce the dimensionless
standard real number $\epsilon$  whose magnitude determines the situations
in which the particle's behaviour is nearly classical.

{\bf Definition 1:} The collimation of a particle through the
slit $]z_1 ,\ z_2[$ is said to be ``$\epsilon$ sharp'' on the open subset
$U$ if the value $Z_{Q}(U)$ of its z-coordinate satisfies both the
following conditions:

\begin{equation}\  Z_{Q}(U) \in  ]z_1,\ z_2[  \end{equation}

\begin{equation}\ z_1\ \leq\ Z_{Q}(U) -\ {s(Z)(U) \over \sqrt{\epsilon}}
\ < Z_{Q}(U) +\ {s(Z)(U) \over \sqrt{\epsilon}} \ \leq\ z_2.
\end{equation}

The inequalities hold pointwise on $U$.

The following theorem follows from algebraic manipulations of the
definitions:

{\bf Theorem 1:}

When the collimation of a particle through the
slit is $\epsilon$ sharp on $U$:

\begin{equation}\ {4s(Z)(\hat \rho)^{2} \over (z_2\ -\ z_1)^2}\
=\ 4{Tr(\hat \rho.\hat Z^2)\ -\
Tr^2(\hat \rho.\hat Z)\over (z_2\ -\ z_1)^2}\ \leq\
\epsilon,\ \forall\ \hat\rho\ \in\ U.\end{equation}

That is,

\begin{equation}\ {4s(Z)(U)^{2} \over (z_2\ -\ z_1)^2}\ \leq\
\epsilon.\end{equation}

Therefore if the slit is narrow then $z_{Q}(U)^{2} - (z)^{2}_{Q}(U)$ 
is small. This means that the quantum real number $Z_{Q}(U)$ can be well
approximated by a constant real number in the interval
$]z_1,\ z_2[$. If the value of the z coordinate of the particle was
measured now it would yield a standard real number in $]z_1,\ z_2[$

When two quantum numbers, corresponding to a position and its conjugate
momentum, are simultaneously $\epsilon$-sharp collimated through slits,
a lower bound on the product of the widths of the slits is obtained.
Since the width of the slit gives a measure of the range of values that
could be obtained if the quantities were measured this represents the
limitation in accuracy imposed by Heisenberg's uncertainty principle.

{\bf Theorem 2}

Let $z$ and $p$ represent the position and conjugate momentum of a
$U$-prepared particle, and let $]z_{1},\ z_{2}[$ and $]p_{1},\ p_{2}[$
be the corresponding slits. If the particle is $\epsilon$-sharp
collimated through both slits then the product of the widths of the
slits must satisfy,

\begin{equation}|z_{2}-z_{1}||p_{2}-p_{1}| \geq \ 2\hbar/ {\epsilon }
\end{equation}

{\bf Proof:}

If the $U$-prepared particle is  $\epsilon$-sharp collimated
through both slits then by Theorem 1,
${4s(Z)(\hat \rho)^2\over (z_2\ -\ z_1)^2}\ \leq\ \epsilon $, for all
$\hat{\rho}\in{U}$, and ${4s(P)(\hat \rho)^2\over (p_2\ -\ p_1)^2}\
\leq\ \epsilon $, for all $\hat{\rho}\in{U}$.
But Heisenberg's inequality states that
${s(Z)(\hat \rho)}\cdot {s(P)(\hat \rho)}
\geq {\hbar/2}$ for all $\hat{\rho}\in{U}$.

Whence $(z_2-z_1)^{2}$. $(p_2-p_1)^{2}\ \geq\ (2\hbar/\epsilon)^{2}$ as
required.
\ \ $\#$

This result determines the minimum area in the classical phase space
that is required if a particle is to be $\epsilon$-sharp collimated
in both the z and p variables.

On the assumption that $]z_1,z_2[ \cap \sigma(\hat Z)
\neq \emptyset$,where $\sigma (\hat Z)$ is the spectrum of $\hat Z$,"
consider $\hat P$, the orthogonal projection operator associated to the
slit $]z_1,z_2[$ via the spectral family for
$\hat Z$,
$\hat P = \hat E_{\hat Z}(]z_1,z_2[)$ and its corresponding quantum
real number  $P_{Q}(U)$ for the open set $U$.

{\bf Theorem 3:}

If the collimation of a $U$-prepared particle through a slit is
$\epsilon$ sharp, then for each $\hat\rho$ in $U$,
\begin{equation}\label{prop3} Tr(\hat P\cdot \hat\rho) \ >\ \ 1\ -\
\epsilon.\end{equation}

In terms of the quantum number $P_{Q}(U)$,
\begin{equation}\ 1\ -\epsilon \ < P_{Q}(U) \leq 1.\end{equation}

{\bf Proof:}

The essential ingredient of the proof is Chebyshev's inequality of
which we shall first recall an elementary derivation.

Let us assume that a random, standard real and positive variable
$z$ obeys a normalised distribution given by the function $\mu(z)$
(we assume that $\mu$ is sufficiently regular so that all the
integrals introduced in the following treatment uniformly converge).
 Let us denote by $<z>$ the average value of the variable $z$:
 $<z>\ =\ \int_o^{\infty} d \mu(z)\cdot z$.

Obviously,  $<z>\ \geq\
\int_{\delta}^{\infty} d\mu(z)\cdot z\ \geq\ \delta \cdot
\int_{\delta}^{\infty} d\mu(z)\ =\ \delta \cdot
\mu(z\,\geq\,\delta).$

Let us consider the variable $(z\,-\,<z>)^2$. By the same reasoning,
we get that:

$\mu((z\,-\,<z>)^2\,\geq\,\delta^2) \,\leq\,{<z^2>\,-\,<z>^2\over
\delta^2}\ =\ {\sigma^2\over \delta^2}$,

where $\sigma$ is the mean
square root deviation of the distribution. As a corollary, we get the
Chebyshev inequality:

${\sigma^2\over \delta^2}\,\leq\,\epsilon$ $\Longrightarrow$
$\mu(|z\,-\,<z>|\,\geq\,\delta) \,\leq\,\epsilon$.

Let us now make use of Chebyshev's inequality for the open set $U$
of $\epsilon$ sharp collimated states. For all $\hat\rho$ in $U$:

 $z_1\ \leq\ Tr(\hat\rho.\hat Z)\ -\ {s(Z)(\hat \rho)\ \over \
\sqrt{\epsilon} }<\
Tr(\hat\rho.\hat Z)\ +\ {s(Z)(\hat \rho)\ \over
\ \sqrt{\epsilon} }\ \leq\ z_2$.

We can identify $s(Z)$ with $\sigma$ provided that we identify the
spectral measure associated to the quantum number $z$ evaluated at
$\hat\rho$ with the measure $\mu$ introduced in the derivation of
Chebyshev's inequality.

Then $Tr(\hat\rho.\hat Z)\ =\ <z>$,
and $\mu(|z\,-\,<z>|\,\geq\,|<z>\ -\ z_1|)\
\leq\ \epsilon$ as
$ {\sigma^2\over (<z>\ -\ z_1)^2}\ \leq\ \epsilon$.

We obtain in a similar way that
$\mu(|z\,-\,<z>|\,\geq\,|z_2\ -\ <z>|)\ \leq\ \epsilon$
as $ {\sigma^2\over (z_2\ -\ <z>)^2}\ \leq\ \epsilon$.

This implies that
$\mu(z\ \notin\ ]z_1,\,z_2[)\ \leq \ \epsilon .$
Therefore in virtue of the
normalisation of the spectral measure,

$Tr(\hat P\cdot
\hat\rho)\ =\ \mu(z_1 < z < z_2)\ >\ 1\ - \ \epsilon$\ \
for all $\hat\rho$ in $U$. $\#$

Intuitively, up to an $\epsilon$, the particle is located
inside the slit. We will reformulate this property in terms of
probabilities in the next section.

 The concept of $\epsilon$ sharp
collimation is tightened by requiring that 
$Tr|(\hat\rho\ - \ \hat P\cdot \hat\rho \cdot \hat P)|\mid \ < \epsilon$
on the open set $U$. Here the absolute value of an operator $\hat A$
is defined by
$\mid \hat A \mid = \sqrt{\hat A^{*} \hat A} $.
This a stronger condition because
$Tr|(\hat\rho\ - \ \hat P\cdot \hat\rho \cdot \hat P)|
\geq Tr( \hat\rho\ - \ \hat
P\cdot \hat\rho \cdot \hat P)$, and $\epsilon$ sharpness on $U$ 
implies that $Tr(\hat P\cdot \hat\rho) \ > ( 1 - \epsilon) $ holds on $U$
which only implies that
 $Tr( \hat\rho\ - \ \hat
P\cdot \hat\rho \cdot \hat P) < \epsilon$ on $U$.  The earlier
theorems remain valid for strict $\epsilon$ sharpness
because the definition requires $\epsilon$ sharpness.

{\bf Definition 1':}

The collimation of a $U$-prepared particle through a slit is
"strictly $\epsilon$  sharp" if it is $\epsilon$ sharp (Definition 1)
and if, for each $\hat\rho$ in $U$,
$Tr|(\hat\rho\ - \ \hat P\cdot \hat\rho \cdot \hat P)|\mid \ < \epsilon$.

{\bf Corollary 1 to Theorem 3:}

Let $\hat\rho_1$ be the restriction of $\hat\rho$
to the slit:
$\hat\rho_1$ = ${\hat P\cdot \hat\rho \cdot
\hat P\over Tr(\hat P\cdot \hat\rho \cdot \hat P)}$.
When $ \rho \in U$, a set of strictly $\epsilon$ sharp collimated
states, the difference between $\hat\rho_1$ and $\hat\rho$ satisfies
$ Tr\mid\hat\rho\ - \ \hat\rho_1\mid \ < \epsilon \cdot {(2 -\epsilon)
\over (1 - \epsilon)}$.

{\bf Proof:}

We have that $ Tr\mid\hat\rho\ - \ \hat\rho_1\mid$ =
$Tr\mid  \hat\rho\ - \ {\hat P\cdot \hat\rho \cdot
\hat P\over Tr(\hat P\cdot \hat\rho \cdot
\hat P)}\mid$ =
$Tr | {(Tr(\hat P\cdot \hat\rho \cdot
\hat P))\cdot \hat\rho\ - \ \hat P\cdot \hat\rho \cdot
\hat P\over Tr(\hat P\cdot \hat\rho \cdot \hat P)}  |$. 
If we put $\mu  = Tr(\hat P\cdot \hat\rho \cdot \hat P)$, this can be
written as $Tr | {\mu  \hat\rho\ - \ \hat P\cdot \hat\rho \cdot
\hat P\over \mu}  |$ = $Tr | {\mu ( \hat\rho\ - \ \hat P\cdot
\hat\rho \cdot \hat P + \ \hat P\cdot \hat\rho \cdot \hat P) -\ \hat P\cdot
\hat\rho \cdot \hat P \over \mu} |$ = $Tr | {\mu ( \hat\rho\ - \ \hat
P\cdot \hat\rho \cdot \hat P) + \ (\mu - 1) (\hat \rho - \hat P\cdot
\hat\rho \cdot \hat P) \over \mu} |$ $\leq$ $Tr | ( \hat\rho\ - \ \hat
P\cdot \hat\rho \cdot \hat P)| + Tr |\hat P\cdot
\hat\rho \cdot \hat P|{(\mu - 1) \over\mu}  $ . Here 
we used the triangle inequality  for the norm $||\hat A||_{1} =
Tr |\hat A| $ on the trace class operators. But $ |\hat P\cdot \hat\rho
\cdot \hat P| = \hat P\cdot \hat\rho \cdot \hat P$ because the latter
is a positive, self adjoint operator, and $Tr|(\hat\rho\ - \ \hat
P\cdot \hat\rho \cdot \hat P)|< \epsilon $ because $ \rho \in U$, 
a set of strictly $\epsilon$ sharp collimated states.
Therefore, 
 $ Tr\mid\hat\rho\ - \ \hat\rho_1\mid < \  \epsilon \cdot 
{(2 -\epsilon) \over (1 - \epsilon)}$

In the last step we used that $ (1 - \epsilon) < \mu = Tr(\hat P\cdot
\hat\rho \cdot \hat P) =  Tr( \hat P\cdot \hat\rho) \leq 1 $ and
${|\mu - 1| \over |\mu|} < {\epsilon\over(1 - \epsilon)}$.   
$\#$

If $Z$ has only one simple eigenvalue in the interval then
$\hat\rho_1$ is the eigenspace projection. 

\subsection{The von Neumann transformation law for quantum numbers.}

When the collimation of a particle through the slit is
strictly $\epsilon$-sharp, many quantum numbers transforms,
up to an $\epsilon$, as if the corresponding operator had undergone a
von Neumann transformation. The von Neumann transformation law for a slit
with an associated projection operator $\hat P$ states that in passing
through the slit any operator $\hat M$ associated with the particle is
changed to $\hat P \hat M \hat P$. The following theorem can be
interpreted as saying that the von Neumann transformation law gives a
good approximation to the quantum real number value of a quantity
associated with a strictly
$\epsilon$-sharp collimated particle.

{\bf Theorem 4:}

If $U$ is an open set of strictly $\epsilon$ sharp collimated states 
for the slit $]z_1,z_2[$, then for all $\hat\rho$ in $U$ and
for all continuous self-adjoint operators $\hat M$,

\begin{equation}|Tr(\hat M\cdot
\hat\rho )\ -\ Tr(\hat P \hat M \hat P\cdot \hat\rho )|\ \leq\ m\cdot
\epsilon\end{equation}

where $m$ is some finite number that depends on $\hat M$. That is, on $U$,
the quantum number $M_{Q}(U)$ is well approximated by the quantum
number $(PMP)_{Q}(U)$
\begin{equation} \ |M_{Q}(U) \ -\ (PMP)_{Q}(U)|\ \leq \
\ m \cdot \epsilon. \end{equation}

{\bf Proof:}

For any operator $\hat M$,

$Tr(\hat M\ -(\hat P \hat M \hat P) \cdot \hat\rho $ =
$Tr((I - \hat P) \hat M (I - \hat P) \cdot \hat\rho )\ +\ \ Tr((I - \hat
P) \hat M \hat P) \cdot \hat\rho )\ +\ Tr( \hat P  \hat M (I - \hat
P) \cdot \hat\rho )$.

(a) If $M$ is a bounded self adjoint operator the following 
estimates hold when $\hat\rho \in U$,

$|Tr(\hat P \hat M (I - \hat P) \cdot \hat\rho )| \leq \
|| \hat M||\cdot Tr |(I - \hat P) \cdot \hat\rho )| < \ 
||\hat M || \cdot \epsilon $,
 
similarly, 
$|Tr((I - \hat P) \hat M \hat P \cdot \hat\rho )| \leq \
|| \hat M||\cdot Tr |\hat\rho \cdot (I - \hat P) |  < \ 
||\hat M || \cdot \epsilon $,
 
and also, 
$|Tr((I - \hat P) \hat M (I - \hat P) | < \ ||\hat M || 
\cdot \epsilon $,
where $||\hat M||$ is the operator norm of $M$.

Therefore, for bounded operators the stated inequality holds 
with $m = 3 ||\hat M ||$.

(b) The result can be extended to a wider class of
quantities associated with unbounded operators
but more care needs to be taken with the topology on
$\Sigma$ with respect to which the operators are continuous
\cite{adelman2}. We show that if
$\hat M$ is the operator $\hat Z$ that defines
the slit $]z_1,z_2[$, then even when $\hat Z$ is unbounded
the result holds provided that

\begin{equation}\label{min} 2\cdot |z_2 \ -\ z_1|\,\leq\,{\epsilon}\cdot
m,\end{equation} where the constant
$m$ = $2\cdot min(|z_1|, |z_2|)$.

For example, if $0\,\leq\,z_1\,\leq\,z_2$, it is easy
to show that, in virtue of Chebyshev's inequality, when $\hat\rho$
is in the set $U$,
$(1\,-\,\epsilon)\cdot z_1\,\leq\,Tr(\hat P\hat Z
\hat P\cdot \hat \rho )\,\leq\,z_2$. For $\epsilon$ sharp
collimation, $Tr(\hat \rho.\hat Z)\ \in\  ]z_1,\
z_2[$ . Therefore,
$|Tr(\hat Z\cdot \hat \rho )\ -\ Tr(\hat P\hat Z \hat P\cdot
 \hat \rho )|\,\leq\,(z_2\,-\,z_1)\,+\,\epsilon\cdot
z_1\,=\,(z_2\,-\,z_1)\,+\,\epsilon\cdot min(|z_1|, |z_2|) \leq
{\epsilon}\cdot m$ when the condition (\ref{min}) is satisfied. $\#$

\subsection{Basic Postulate 1.}

The usual frequency concept of probability is implicitly present
in our description of the one slit experiment when applied to an
ensemble of particles. Sometimes a particle passes through the slit and
sometimes it is stopped at the barrier because each particle
always has a position given by a quantum real number. When the
$U$-prepared particles are all $\epsilon$ sharp collimated and if
$\epsilon$ approaches zero then
$s(Z)(U)$ becomes infinitely smaller than the extension of the slit so
the quantum particle should behave like a pointlike particle and pass
through the slit with probability one. That is, $\epsilon$ sharp
collimated quantum numbers approximate, to within $\epsilon $,
standard classical numbers in the spectrum of the self-adjoint operator
associated with the quantity being collimated. This suggests that
the measurable values of a quantum number must belong to
the spectrum of the self-adjoint operator.

These considerations recall the philosophy of Niels
Bohr in which the measurement process gives an interface between the
classical and quantum worlds. In this interface quantum potentialities
become actual, a process sometimes called the  objectification process.
In terms of quantum real numbers, it is a process in which the
quantum real numbers become sharp and realise standard, classical values.

{\bf Basic Postulate 1:}

(i) The measured values of physical quantities always belong to the
spectrum of the corresponding self-adjoint operator. 
(ii) If a quantity is associated with the self-adjoint operator $\hat M$
 and if $U$ is an open subset of strictly $\epsilon$ sharp collimated
states for the interval $]m_1,m_2[$ in the spectrum of $\hat M$ , then the
probability that the quantum number $M_{Q}(U)$  belongs to
$]m_1,m_2[$ is larger than $1-\epsilon$.

Note that by Theorem 3, if $\hat P = \hat E_{\hat M}(]m_1,m_2[)$ is
the projection operator for the interval $]m_1,m_2[$ then we can
identify $P_{Q}(U) > 1-\epsilon$ as the probability of passing through
the slit when $U$ is an open subset of strongly $\epsilon$ sharp
collimated states for the interval.

There is a special case,  when the associated operator
$M$ has only one eigenvalue
$\lambda$ in $]m_1,m_2[$ so that $\hat P$ is the projection
onto the eigenspace for $\lambda$ and if $U$ is an open set of strongly 
$\epsilon$ sharp collimated states for $]m_1,m_2[$, 
then the probability that
$M_{Q}(U) = \lambda$ is $P_{Q}(U) > 1 - \epsilon$.

Except in the special case described above there will be many
different standard real numbers in the interval that could be realised
as the measured value of of the quantity. In a measurement the quantum
real number value $M_{Q}(U)$ of the quantity is forced to realise  
one of them. We do not know with what probability the different sharp
values will occur. Basic postulate 1 gives an estimation of the
probability of passage through the slit only in the nearly deterministic
regime, i.e., in situations of sharp collimation. In the next section we
start to evaluate the probabilities of getting different outcomes in 
simple cases.

\section{Deduction of the quantum probability law.}
 
In the standard quantum mechanics literature the question of whether
quantum probability rules like Born's should be postulated independently
has been discussed. Several attempts to derive the quantum probability
rule by considering many copies of a system and postulating the validity
of the eigenstate rule have been made \cite{squires}. The eigenstate
rule states that if the system was prepared in an eigenstate of a
self-adjoint operator $\hat A$ then any subsequent measurement of
$\hat A$ always yields the corresponding eigenvalue.

The eigenstate rule is not equivalent to our first basic
postulate. Nevertheless, our first basic postulate implies an
modified version of the eigenstate rule in the special case when $\hat A$
has only one eigenvalue $\lambda$ in the interval $]a_1,a_2[$. 

If we are content with equality up to an $\epsilon$ we can show that
the basic postulate 1 implies Born's rule. We
shall only sketch the proof in the case  of the simplest quantum
experiment, a dichotomic experiment,  following the treatment given by
Goldstone \cite{goldstone} and  modified by Squires \cite{squires}. A
general proof of a similar result was obtained by Busch for observables
with a continuous spectrum \cite{busch}.

We require another postulate, the "ergodic assumption"

{\bf Basic Postulate 2}
The result of an average measurement performed at the same time on $N$
identical copies of a system and the averaged result of $N$ individual
measurements performed successively in time on $N$ identically prepared
systems are identically distributed.

{\bf Theorem 5:}
If basic postulates 1 and 2 are satisfied,
and the system is prepared in an neighbourhood
$W$ of the density matrix $\hat\rho_0$,
$W = \{\rho| Tr\big|(\hat\rho \ -\ \hat\rho_0) \big|$ $\ <\
\epsilon\}$
then, up to $\epsilon$, the
probability of measuring the outcome $i$, for $i=1,0$, equals
$Tr \hat\rho_0 \hat P(i)$, where $\hat P(1)$ is the projection operator
of the slit and $\hat P(0)\,=\,1\,-\,\hat
P(1)$. That is,
\begin{equation}|P_{Q}(i)(W)\ -\ Tr(\hat\rho_0\hat P(i))|\ 
< \epsilon.\end{equation}

{\bf Proof:}

Following the notation introduced by Squires \cite{squires}, we define
an ``average'' operator $\hat Q$ constructed to give the average value 
of a dichotomic quantity, with values 1 or 0, associated to the
passage through the slit.  On account of this choice of values for the
quantity, the average value can be identified with the relative
frequency of passage through the slit.

$\hat Q$ =
${1\over N}\sum_{i:\ 1...N}\ \hat Q_{i}(1)$,
where $\hat Q_{i}(1) = \bigotimes\hat I_1\bigotimes ....\bigotimes\hat
P_{i}(1) \bigotimes ...\bigotimes\hat I_N$, an N-fold tensor product with
the identity operator in each slot except the
$i^{th}$ which contains $\hat P_{i}(1)$. It is easy to check that the
spectrum of $\hat Q$
goes from 0 to 1 by steps ${1\over N}$. This is due to the fact that the
spectrum of each
projector $\hat P_{i}(1)$ is equal to $\{0,1\}$.

We consider $N$ identical copies of the density matrix 
state $\hat\rho_0$,
$\hat \Omega_0 = \bigotimes_{j=1}^{N}\hat\rho_0(j)$, 
of each state $\hat \rho \in W$, $\hat \Omega =
\bigotimes_{j=1}^{N}\hat\rho(j)$ and 
of the open set $W$, 
$\bar W = \prod_{j=1}^{N} W(j)$.

Now $Tr(\hat \Omega_{0} \cdot \hat Q) = Tr(\hat \rho_{0} \cdot \hat P(1)),
\neq 0$ by assumption, $Tr(\hat \Omega \cdot \hat Q) = Tr(\hat
\rho \cdot \hat P(1))$ and hence $Q_{Q}(\bar W) = P(1)_{Q}(W) = p(W)$.

Now consider the N-particle projection operators that are a
tensor product of $J$ ``yes'' single particle projections , 
$\hat P_{i}(1)$,  and $(N - J)$ ``no'' single particle projections,
$\hat P_{i}(0)$. By permuting the order
of the single particle operators we deduce that for each $J$ there are
${N\choose J}\,=\,{N!\over J!\ (N\,-\,J)!}$ such N-particle operators
which represent N-particle measurements in which $J$ particles do, and
$(N-J)$ do not, pass through the slit.
 For any of these N-particle projections $\hat Q_J$ and any $\hat \rho
\in W$, 

$Tr(\hat \rho \cdot \hat Q_J) = $
 $(Tr(\hat \rho \cdot \hat P(1)))^{J}$ $(Tr(\hat \rho \cdot \hat
P(0)))^{(N-J)}.$

Therefore if prepared in the open set $W$ the quantum real number
associated to the
situation in which $J$ particles do and
$(N-J)$ don't pass through the slit is 
${N\choose J} \cdot (P(1)_Q(W))^{J}$ $(P(0)_Q(W))^{(N-J)}.$ 

 Now, this is just the expression for the probability of having $J$
favourable and $N-J$ unfavourable events in a Bernouilli process 
with probabilities $(P(1)_Q(W))$ and $(P(0)_Q(W))$. 
This is only a formal identification but the expression can be manipulated
mathematically. 

In the limit as $N$ goes to infinity the Bernouilli (binomial)
distribution can
be approximated by a Gaussian distribution:

 $\,{N\choose J}  \cdot p(W)^{J}\
q(W)^{N\,-\,J}$
$\sim\,{1\over\sqrt{2\pi\,N \,p(W)\, q(W)}}
\cdot $exp$ -( J\,-\,Np(W))^2
\over 2\,N\,p(W)\,q(W)$.

Also, the spectrum of $\hat Q$ tends to cover the unit interval in this
limit and  this Gaussian distribution scales to a normal density function 
for the relative frequency $x = J/N$ given by
$\psi (x)$ = $(2\pi p(W) q(W)/N)^{-1/2}\cdot exp
-[(x - p(W))^{2}/(2p(W)q(W)/N)]$. Therefore
$\psi$ has mean $<x> = p(W)$ and standard deviation $(p(W)q(W)/N)^{1/2}$.

In virtue of Chebyshev's
inequality,
$\mu(|x\,-\,<x>|\,\geq\,\delta) \,\leq\,{(p(W)q(W)/N)\over\delta^2}$.

Therefore, since $\psi(x) $ is the probability density function for the
relative frequency $x$,
$\mu(|x\,-\,<x>|\,\geq\,\delta)= 1 - \int_{<x>-\delta}^{<x>+\delta}\psi
(x) dx$ . On using  $\delta\ =\ {1\over
\sqrt{ N^{1\ -\ \lambda}}}$, with $\lambda$ a 
positive standard real number between 0 and 1, the probability of the
frequency lying in the interval [$p(W) - {1\over \sqrt{ N^{1 -
\lambda}}},\ p(W) + {1\over \sqrt{ N^{1 - \lambda}}}$] is larger than 
1 - $ p(W) \cdot q(W)\over N^{\lambda}$ which shows how the relative
frequency approaches $p(W)$ in the large $N$ limit.

Suppose that we measure the relative frequency, 
represented by $\hat Q$, with a measuring device of resolution $2R$, 
which means that we are unable to distinguish values that
are less than a distance $2R$ apart. For any realistic device, 
$R$ can be assumed to be a very small standard real number but is never
equal to zero. By choosing $N$ sufficiently large, we can always
ensure that the standard deviation $S(Q)(\bar W) = (p(W)q(W)/N)^{1/2}$ is
much smaller than the resolution $2R$ of the apparatus. Actually,
whenever $N$ is larger than ${ p(W)\cdot q(W)\over \epsilon \cdot R^2} $
and then $\bar W$ is a set of $\epsilon$ sharp collimated states for the
slit ]$p(W)\ -\ R,\ p(W)\ +\ R$[ for measuring $\hat Q$.  
Therefore by basic postulate 1,for systems prepared in $\bar W$ the
probability that we observe values of $Q_{Q}(\bar W)$ that belong to the
interval  ]$p(W)\ -\ R,\ p(W)\ +\ R$[ is 1 to within an $\epsilon$. This
means that, with probability 1 up to an $\epsilon$, the result of the
measurement of
$\hat Q$ will be equal to $p(W)$ to within the resolution
$2R$ of the apparatus.

Now, by basic postulate 2, the averaged
result of $N$ individual dichotomic measurements performed
successively on $N$ identically prepared systems in the
open set $W$ and the results of an average
measurement of $\hat Q$ performed at the same time on
$N$ identical copies of a system in the open set $\bar W$ are equally
distributed. Since the result of measuring
$\hat Q$ is almost certainly $p(W)$ in the sense made
precise before, we have that in the limit of
large $N$ the average value of the individual operator
$\hat P_i(1)$ is certainly equal to $p(W)$.
This average value obtained after $N$ individual
measurements is precisely the frequency or probability
of obtaining the result "yes" in an individual
measurement. 

Finally we must show that this result holds uniformly
over $W$. Each density matrix $\hat\rho$ in $W$
satisfies $Tr|(\hat\rho \ -\ \hat\rho_0)|$ $\ <\ \epsilon$.  
But, for $i = 0, 1$,$\hat P(i)$ is a bounded operator of norm 1,
therefore,
$|Tr (\hat\rho\hat P(i))\ -\ Tr(\hat\rho_0\hat P(i))|\ <\ \epsilon $. 

That is, for each $i = 1,0$, the probability of measuring
the outcome $i$ is essentially constant, up to an $\epsilon$,
and equal to $Tr(\hat\rho_{0}\hat P(i))$
for all density matrices in the neighbourhood $W$. In terms
of the quantum numbers $P(i)_{Q}(W)$,$i = 1,0$,
$|P(i)_{Q}(W) - \gamma(i)|\ <\ \epsilon$
where the standard real number $\gamma(i) =
Tr(\hat\rho_{0}\hat P(i))$.

This completes the proof that the basic postulates 1 and 2 are
sufficient to derive Born's quantum probability rule because when 
$\hat P(1))$ projects 
onto the 1 dimensional space spanned by $\big|1 \big>$ and
the state $\hat\rho_0$ is pure and equals 
$\big|\psi \big>\big<\psi\big|$ where $\psi$  is a unit vector
then  $Tr(\hat\rho_{0}\hat P(1))$ equals
the Born rule expression,
$\mid\big<\psi\big|1 \big>\mid^2$. $\#$

The generalisation of this argument to a finite sequence of dichotomic
observations and thus to an arbitrary discretised measurement process is
straightforward. If any realistic experiment can only have a finite
number of outcomes then we have established the frequency meaning of
probability for realistic experiments. We have still to develop 
dynamical models of how the different outcomes are realised.

It is easy to show that, in virtue of the Theorems 3 and 5,
when the collimation of a particle through the slit
is $\epsilon$ sharp, the particle will pass through the
slit with probability equal to 1 (up to an $\epsilon$). This
shows the internal consistency of our choice of axioms.

\section{Persistence of measured values.}

Let us return to the single slit experiment as the prototype of a 
class of measurements in which the measured values persist. In
order to guarantee the persistence  of the observed values of the
positions of the particle, we must impose  the following continuity
condition: immediately after the particle  has passed through the slit,
the probability is negligible of finding it elsewhere than in the
vicinity of the slit. That is, there exists a standard real number $0 <
\epsilon << 1$ and an open set $U$ such that for each $\hat\rho$ in $U$,
\begin{equation}
\label{prop3} Tr(\hat P\cdot \hat\rho) \ >\ \ 1\ -\
\epsilon,\end{equation} 
$\hat P$ being the spectral projection for the slit.

{\bf Basic Postulate 3 A:}

If a quantity $\hat A$ is measured and found to have values in the
subset $I$, then there exists a standard real number $0 < \epsilon << 1$
such that immediately after the measurement, the system belongs to the
largest open set $U$ on which
$Tr|(I - \hat P)\cdot
\hat\rho| \ <\ 
\epsilon $,  $\hat P$ being the spectral projector of $\hat A$ onto $I$.

Consequently, in terms of the quantum number $P_{Q}(U)$,
\begin{equation}\ 1\ -\epsilon \ < P_{Q}(U) \leq 1.\end{equation}

{\bf Proposition 1:}

When the basic postulates 1, 2 and 3 A are satisfied, if a subset $I$
of values of a quantity is measured, then, just after the
measurement, the quantum numerical value of the quantity will
still belong to $I$ with probability close to one. 

{\bf Proof:}

Let $\hat A$ denote the self-adjoint operator of 
to the quantity being measured and let us assume that the measured values 
belong to the subset $I$.
The basic postulate 3 A implies that, if $\hat P$ is the spectral
projection operator for $\hat A$ on the subset $I$, then immediately
after the measurement the system is in the set $U$ on which 
$1\ -\epsilon \ < P_{Q}(U) \leq 1$. By a straightforward
application of the Born rule, the validity of which was established in
theorem 5, this means that the probability that the quantum number
$A_R(U)$ belongs to $I$ is greater than $1\ -\epsilon$.$\#$

We further note that in the limit of vanishing
$\epsilon$, any bounded observable $\hat B$ transforms according
to the von Neumann transformation rule on the open set $U$ because for
$\rho \in U$,

$\mid Tr(\hat\rho\cdot\hat
B)\ -\ Tr(\hat\rho\cdot\hat P\hat
B\hat P)\mid\,\leq\,||B||\cdot\epsilon$.

The proof of this result is the same as
that of Theorem 4 which dealt with the case
of strictly $\epsilon$ sharp collimated states.

{\bf Comment:}

The persistence/continuity in time of the quantum real number values of
the particle's position was implicitly assumed when we described a passage
of a particle through a slit in the first section. The persistence, up
to an
$\epsilon$, of the measured values of  a particle's position is based
upon experimental facts,  exhibited in the setting up of sources and
targets and in bubble  chamber pictures when one sees a temporal sequence
of aligned  excitations, which approximate, up to an $\epsilon$, 
classical  continuous trajectories that exhibit the persistence of
localisation.

\subsection{The preparation process as a filtering process.}

Note that basic postulate 3A is necessary in
order to establish the relevance of basic postulate 1
and of Theorem 5. In order that a state is $\epsilon$ sharp collimated,
the particle must be  physically prepared. Similar preparation of $N$
copies of an open set $W$ is needed in the derivation of the theorem 5.
This can be done, in principle, using a combination of 
dynamical evolutions (that we shall describe in a next section) and
filtering processes.

 Suppose that during the preparation
process different quantities are measured successively. It
is well-known that if the quantities are represented by commuting
operators, the Birkhoff-von Neumann lattice of physical properties admits
a classical (Boolean) representation \cite{birkhoff1}; this suggests that
classical logic describes the logic of the outcomes (up to
$\epsilon$). In the standard theory this Boolean
representation does not exist for quantities represented
by non-commuting operators because the distributivity
property of the lattice is violated \cite{birkhoff1}. A similar
conclusion is obtained in axiomatic probability theory, the
violation of Bell's inequalities can be shown to reflect the
non-existence of a classical probabilistic structure
underlying quantum probability \cite{gutko}. However in  
the quantum real number model the logic is intuitionistic
\cite{adelman1}. If the outcomes of the measurements are given by quantum
real number values then, as Theorem 2 shows, if limited  accuracy is
accepted, quantities represented by non-commuting  operators can be
measured in succession. The  logic of propositions is then
intuitionistic but not Boolean  in general.To ensure that Boolean logic 
holds more conditions  have to be imposed on the measurements. We will
not pursue this discussion further in this paper. Nevertheless, for the
registered outcomes of measurements, classical, Boolean logic and
probability  rules are relevant. Note that, in last resort, it is only
through the development  of dynamical models of the measurement process
that it ought to be possible to connect the quantum and the classical
worlds.

In standard quantum mechanics, when observables commute, the temporal
order in which they are measured does not affect the statistical
distribution of the outcomes\footnote{For instance, it can be shown that
when the system is
an entangled bipartite system of which the components belong to regions of
space-time separated by a Minkoskian spacelike vector, the quantum
statistical correlations
between both systems are the same as when these regions are separated by a
timelike vector.
In the latter case, the chronology of the measurements is invariant under a
Lorentz
transformation. Otherwise, the temporal order depends on which inertial
referential is
chosen in order to describe the experiment \cite{gisin}. }.Therefore,
the outcome observed during the measurement of a quantum number
$A_Q$ should persist when another number $B_Q$ is
measured provided $\hat A$ and $\hat B$ commute.
This discussion is encapsulated in the following postulate:

{\bf Basic Postulate 3 B:}

Suppose one quantity is measured and found to have values 
in the subset $I$ and directly afterwards a second quantity is 
measured. If the quantities are represented by strongly commuting
operators $\hat A$ and $\hat B$ and if $\hat P$ is the spectral 
projector of $\hat A$ onto $I$, then, just after the measurement 
of $\hat B$, the system still belongs to the open set $U$ on which
$Tr|(I - \hat P)\cdot \hat\rho| \ <\ \epsilon$.  Moreover, the temporal
order in which $\hat A$ and $\hat B$ are measured does not  affect the
statistical distribution of the outcomes.

Note that an alternative approach was proposed elsewhere
\cite{ballentine} in order to describe the joint-measurement of the
observables $\hat A$ and $\hat B$, that is known in the litterature as the
statistical interpretation. The basic idea is that, being considered that
the temporal ordering of the measurement of $\hat A$ and $\hat B$ does
not matter, it is sufficient to consider the global measurement as a
whole and to apply the Born rule without considering the possibility of
the collapse of the wave function during partial measurements. Logically,
this is a consistent approach but according to us it does not answer to
the question of the collapse of the full wave function during the global
measurement. It also does not explain why regitered outcomes are
persistent. The concept of persistence introduced by us in the
previous postulates reflects our personal philosophical preference
according to which a measurement is a real process. Both views are
consistent, exactly in the same way that the violation of local realism
by quantum entangled systems can be interpreted as the refutation either
of realism or of locality.

{\bf Consequence of the Basic Postulates 3 A and B:}

If a subset $I$ of values of a quantity $\hat A$ is measured,
and that, directly afterwards, a subset $J$ of values of a
quantity $\hat B$ is measured, and that $\hat A$ and $\hat
B$ strongly commute, then, just after the measurement of $\hat B$, the
system will belong to an open neighbourhood $U\ \cap\ V$,
with $U = \{\rho : Tr|(I - \hat P)\cdot \hat\rho| \ <\ \epsilon\}$
and $V = \{\rho : Tr|(I - \hat P')\cdot \hat\rho| \ <\ \epsilon\}$
where $\hat P$ is the spectral projector
of $\hat A$ onto $I$ and $\hat P'$ is the
spectral projector of $\hat B$ onto $J$.

From the standard quantum mechanics viewpoint,this looks like the
conjunction of propositions being represented, in Boolean logic, by the
intersection of the characteristic sets of the propositions\footnote{This
analogy with classical logics in the case of commuting observables is
also valid for what concerns the logical implication, which corresponds
to the set-theoretical inclusion relation in Boolean
representations. For instance, it is easy to deduce from the definition
1 that, when a system is
$\epsilon$ sharp collimated relatively to a slit of breadth
$z_2 \ -\ z_1$, it will certainly (up to an $\epsilon$) pass
through a parallel and non-distant larger slit of breadth
$\tilde z_2 \ -\ \tilde z_1$ (with $\tilde z_2 \ -\ \tilde
z_1\ >\ z_2 \ -\ z_1$) the center of which
is aligned with the center of the first slit.}. However the sets
are open, because the logic is intuitionistic \cite{adelman1}.

As a consequence of postulates 3 A and 3B, we can use the language of
quantum numbers to describe preparation processes as sequences of
filtering processes performed on a particle. The prepared state of the
particle is then defined by the set of intervals of quantum real
numerical values of the filtered quantities. A new concept of quantum
state is derived from this set of intervals.  Instead of
claiming that a certain state, represented by a density  matrix, was
prepared, we say that the system underwent a preparation
procedure during which certain quantum numbers were prepared. This 
provides us with an operational definition of the state of a quantum
system in terms of quantum numbers.

The equivalence between preparation and measurement for the
class of processes in which measured values persist allows the passage
from the standard interpretation of quantum theory to that
of quantum real numbers in which physical quantities always have quantum
numerical values that exist to extents given by open subsets of
$\Sigma$. However it is only when the measured quantum real numbers
approximate standard real numbers closely, that is, when they are
$\epsilon $ sharp collimated, and persist, that they become concrete,
recordable facts.

The fact that the observed outcomes of a
measurement persist makes it possible to define more accurately the
transformation undergone by the quantum real numbers during
the measurement process.

\subsection{The Luders-von Neumann transformation rule.}

In the standard quantum theory, the collapse hypothesis is often
given as an independent postulate governing the behaviour of systems
under measurement. It states that if the system  was prepared 
as the density matrix $\hat \rho_0$,then during the measurement $\hat
\rho_0$ "collapses" to $\rho_{0}' = {\hat P\cdot\hat\rho_0\cdot \hat
P\over Tr(\hat P\cdot\hat \rho_0)}$, where $\hat P$ is the projection
operator of the slit. Then any observable
$\hat B$ transforms, in the Heisenberg picture, according to the
Luders-von Neumann transformation rule, in which $\hat B$ is
changed to ${\hat P \cdot\hat B\cdot \hat P\over Tr(\hat P \cdot
\hat \rho_0)}$.

This transformation rule differs from the von Neumann
rule which says that in similar circumstances
$\hat B$ is changed to $\hat P\cdot\hat
B\cdot\hat P$. The difference is due to 
the fact that for the von Neumann transformation the 
preparation of the initial state of the particle includes 
the process of collimation through the slit, while the 
preparation of the initial state for the Luders-von Neumann 
transformation does not. This distinction is emphasised in the two next
propositions.

{\bf Proposition 2 ( Luders-von Neumann rule)}

Assume that a system is initially prepared in the open set
$W$ of states centered on the state $\hat\rho_0$:
$W\,=\,\{\hat \rho\,\in\,\Sigma:\,Tr \big| (\hat\rho \
-\ \hat\rho_0 ) \big|\ <\ \delta\}.$ 
Next assume that during the preparation of the initial state the quantity 
$\hat A$ is measured/prepared with  values in the interval $I$. Then any
quantity associated with a bounded self-adjoint operator$\hat B$  has a
quantum real number value given approximately  by the constant standard
real number 
$ Tr \hat \rho_{0}' \cdot \hat B $, where
$\hat \rho_0'$ =
${\hat P\cdot\hat\rho_0\cdot \hat P\over
Tr(\hat\rho_0\cdot P)}$.

Here $\hat P$ is the spectral projection operator for $\hat A$ on the
interval $I$ and $\hat P\cdot\hat\rho_0 \ne \hat 0$.
The approximation is governed by the preparation
parameter $\delta$ and the persistence parameter $\epsilon$. 

{\bf Proof:}

After the measurement of
the quantity $\hat A$, the system belongs to an open set $U$ defined
in the basic postulate 3 A, on which  $Tr| (\hat I-\hat P) \cdot \hat \rho
| < 
\epsilon$, where
$\hat P$ is the spectral projection of $\hat A$ onto $I$. By assumption,
the initial preparation is also described by the open set $W$ so that for
any
$\hat
\rho
\in W \cap U$,

\begin{equation} Tr |(\hat \rho -\ \hat \rho_{0}' )|
= Tr|(\hat \rho -\ \hat \rho_0 +\ \hat \rho_0 -\ \hat \rho_{0}')|
\leq Tr|(\hat \rho -\ \hat \rho_0)| + Tr|(\hat \rho_0 -\ \hat \rho_{0}')|
\end{equation}

The first term is less than $\delta$ because $\hat \rho \in W$. 

The second is less than $\epsilon \cdot (2 - \epsilon)/(1 - \epsilon)$
by Corollary 1 to Theorem 3 (the proof of which is still valid under the
present assumptions).

Thus
\begin{equation} Tr |(\hat \rho -\ \hat \rho_{0}' )|
\ <\  \delta + \epsilon \cdot (2 - \epsilon)/(1 - \epsilon).
\end{equation} 

Therefore if $\epsilon$ and $\delta$ are small enough
the measured value of the quantity with bounded operator $\hat B$ will
be given to a good approximation by the constant number
$ Tr \hat \rho_{0}' \cdot \hat B $, as predicted by the Luders-von Neumann
transformation. $\#$

Note that in the previous proposition, we assumed that a particular value
of the first quantity $\hat A $ was measured during the preparation
process. The next proposition establishes the Luders-von Neumann rule when
the preparation process is assumed to end before the measurement of the
first quantity $\hat A $.

{\bf Proposition 3 (extended Luders-von Neumann rule)}

Suppose the system has been prepared initially in an open set $W$ of
extension $\epsilon$ around the density matrix $\hat\rho_0$. If 
$\hat A$ is then measured, found to have  values in the interval
$I(i)$,$i = 1,....,N$,and if $\hat P_i$, the spectral projection of $A$
for $I(i)$, satisfies $\hat P_i \cdot \hat \rho_{0} \ne 0$,  then
immediately after the measurement, the system will belong to an open set
$W'(i)$ of extension  $\epsilon$ around the density matrix $\rho_{0}'(i)
= {\hat P_i\cdot\hat\rho_0\cdot \hat P_i\over Tr(\hat P_i\cdot\hat
\rho_0)}$ .

Accordingly, the quantum real number associated to any bounded
self-adjoint operator $\hat B$ that strongly commutes with $\hat A$
transforms as follows: $ B_{Q}(W) -->  B_{Q}(W'(i))$

{\bf Proof of the Proposition 3.}

Consider two strongly commuting self-adjoint operators $\hat A$ and
$\hat B$. To simplify the notation,  we suppose that both operators
are bounded and that $\hat A\ =\ \Sigma_{i=1}^N\,a_i\cdot \hat P_i$ where
$a_i\,\in\, {\bf R},\ N\,<\,\infty$, and $\hat P_i$ is the projection
onto the eigenvalue $a_i$.  Given postulate 3 B the outcomes of 
measurements performed on $\hat A$ persist during measurements of 
$\hat B$ when $\hat A$ and $\hat B$ commute so that we can decompose the
measurement of $\hat A\cdot \hat B$ into the measurement of
$\hat A$ alone followed by the measurement of $\hat B$. In virtue of the
last part of the postulate 3 B, the probabilistic predictions that we
derive from the Born rule will be the same whether we measure the
quantity  $ \hat A \cdot \hat B$ as a single quantity or we measure 
$\hat A$ and $\hat B$ sequentially. 

Let the open set $W\,=\,\{\hat \rho\,\in\,\Sigma:\,Tr\big| \hat\rho \ -\
\hat\rho_0 \big|\ <\ \epsilon\}$ be given, where $\hat \rho_0$ satisfies
\begin{equation}
\hat \rho_0 \cdot \hat P_i \ne 0,
\end{equation} 
for each $\hat P_i$ in the spectral decomposition
of $\hat A$.

For any $\hat \rho\,\in\,W$,

\begin{equation}Tr(\hat\rho\cdot\hat A\cdot\hat B) =
Tr(\hat\rho\cdot\Sigma_{i=1}^N\,a_i\cdot \hat P_i\cdot \hat B) =
\Sigma_{i=1}^N\,a_i\cdot Tr(\hat\rho\cdot \hat P_i\cdot \hat B)
\end{equation}
 But
\begin{equation}|Tr(\hat\rho\cdot \hat P_i\cdot \hat B) -
Tr(\hat\rho\cdot \hat P_i\cdot \hat B \cdot \hat P_i)| \,=\,0.
\end{equation} because $\hat A$ and $\hat B$ strongly commute, so that

\begin{equation}|Tr(\hat\rho\cdot\hat A\cdot\hat B) -
\Sigma_{i=1}^N\,a_i\cdot Tr(\hat\rho\cdot \hat P_i\cdot \hat B\cdot 
\hat P_i)| \,=\,0
\end{equation}

Now since $\hat \rho_0$ was chosen so that
 $\hat P_{i}\cdot \hat \rho_0\,\not=\,0$ for any $i$,

\begin{equation}
\Sigma_{i=1}^N\,a_i\cdot Tr(\hat\rho\cdot \hat P_i\cdot \hat B\cdot
\hat P_i)\,=\,\Sigma_{i=1}^N\,a_i\cdot Tr(\hat\rho_0\cdot \hat P_i)\cdot
{Tr(\hat\rho\cdot
\hat P_i\cdot \hat B\cdot \hat P_i)\over Tr(\hat\rho_0\cdot \hat P_i)}
\end{equation}
for all $\rho\,\in\,W$.
By Theorem 5,
for all $\rho\,\in\,W$,
$Tr|(\hat\rho\cdot
\hat P_i)\,-\ (\hat\rho_0\cdot \hat P_i)|\,<\,\epsilon$ so that
$|Tr(\hat\rho\cdot
\hat P_i\cdot \hat B\cdot \hat P_i)\,-\,Tr(\hat\rho_0\cdot \hat P_i\cdot
\hat B\cdot
\hat P_i)|\,<\,\epsilon\cdot ||\hat B||$. Thus to within an error that
goes to 0 with $\epsilon$, for all $\rho\,\in\,W$,
$Tr(\hat\rho\cdot\hat A\cdot\hat
B)\,=\,\Sigma_{i=1}^N\,a_i\cdot Tr(\hat\rho_0\cdot \hat P_i)\cdot
{Tr(\hat\rho_0\cdot \hat P_i\cdot \hat B\cdot \hat P_i)\over
Tr(\hat\rho_0\cdot
\hat P_i)}$.

Theorem 5 permits us to approximate the term $Tr(\hat\rho_0\cdot \hat
P_i)$ by $(P_{i})_{Q}(W)$. 
We now use the permutation property of the trace to rewrite
${Tr(\hat\rho_0\cdot \hat P_i\cdot \hat B\cdot \hat P_i)\over
Tr(\hat\rho_0\cdot \hat P_i)}$ =
$Tr\hat \rho_0'(i)\cdot \hat B$, where $\hat \rho_0'(i)$ =
${\hat P_i\cdot\hat\rho_0\cdot \hat P_i\over
Tr(\hat\rho_0\cdot \hat P_i)}$  is the Luders-von Neumann transformed of
$\hat \rho_0$ when the outcome $a_i$ has been measured.Thus, for all $
\rho \in W$, $Tr(\hat\rho\cdot\hat A\cdot\hat B)$ is well approximated by
a sum that is independent of $\rho \in W$. By basic postulate 3B the
statistical distribution of the outcomes $a_i \cdot b_j$ is independent
of whether the quantities of $A$ and $B$ were measured simultaneously  or
sequentially. 

This result can be written in terms of quantum numbers,
if $W= \{ \rho; Tr| \rho - \rho_{0} | <  \epsilon \}$ and
$W'(i) = \{ \rho; Tr| \rho - \rho_{0}'(i) | <  \epsilon \}$
where
$\rho_{0}'(i) = {\hat P_{i}\cdot\hat\rho_0\cdot \hat
P_{i}\over Tr(\hat P_{i}\cdot\hat \rho_0)}$,then

\begin{equation}
\ (A\cdot B)_{Q}(W) \,\approx \Sigma_{i=1}^N\, A_{Q}(W'(i))
\cdot(P_{i})_{Q}(W) B_{Q}(W'(i))
\end{equation} That is, the value of $(A \cdot B)_Q$ at $W$ equals the
sum over $i$ of the products of the values of $A_Q$ and $B_Q$ at $W'(i)$
weighted by the probability $(P_i)_Q$ at $W$ to an approximation that
depends on the precision of the initial preparation. Note that this
result is valid in general, even when $\hat P_i \cdot \hat \rho_{0} =
0$ for some $i$.  
$\#$

{\bf Remark}:

 In the quantum numbers interpretation, 
we claim on the basis of the results of the previous subsection that if
the measurement of the quantum number
$A_{Q}$ involves filtering through a slit $I(i)$ then any quantity whose
corresponding operator commutes with $\hat A$ behaves, up to an
$\epsilon$, as if it had undergone a Luders-von Neumann transformation.
Note that this does not imply that the collapse process 
really occurs, but rather that the collapse postulate gives a good
approximation to the quantum numbers obtained in this type of
measurement. Nevertheless, the change
undergone during the measurement process cannot be described solely by a
unitary evolution (this is the core of the so-called measurement problem)
as shows the following example.

\subsection{An example of measurement of position}

The following example shows that when a particle is sharply localised in
space, and that a pointer interacts with this particle according to a well
chosen interaction (in this case an impulsive von Neumann
interaction Hamiltonian), the pointer reveals unambiguously the position
of the particle. In this example, the apparatus being
located in classical space time can only register (reveal)
unambiguously numbers that are approximately classical.

Assume that the system is represented by particle 1, the measuring
apparatus by particle 2. They will be treated as quantum
systems with associated Hibert spaces $\mathcal{H}(1)$ and 
$\mathcal{H}(2)$, while the combined two particle system has the
tensor product Hilbert space 
$\mathcal{H}(1,2)$ = $\mathcal{H}(1) \otimes \mathcal{H}(2)$. The
corresponding state spaces are $\Sigma(1)$, $\Sigma(2)$ and $\Sigma(1,2)$.
When $W(1)$ is an open set in $\Sigma(1)$ and $W(2)$ is open in
$\Sigma(2)$,  we define the superset $W(1,2)$ of $W(1)$ and $W(2)$ to be
the smallest open set in $\Sigma(1,2)$ such the partial traces
$Tr_{\mathcal{H}(1)}\rho(1,2) \in W(2)$ and 
$Tr_{\mathcal{H}(2)}\rho(1,2) \in W(1)$ for all $\rho(1,2) \in W(1,2)$.
  
Initially particle 1 is prepared so that the quantum real number
value of its position is $X(1)_Q(W(1))$ where 
$W(1) \supset U(1)\cup V(1)$. The open sets $U(1)$ and $V(1)$ are 
such that the quantum real number values $X(1)_Q(U(1))$ and
$X(1)_Q(V(1))$ of the particle's position make it $\epsilon$ sharp
collimated in one of the two slits in the screen. If the slits are
determined by the classical numbers $a < b < c < d$ as $I_1 = ]a,b[$ and
$I_2 = ]c,d[$ then $a < X(1)_Q(U(1)) < b < c < X(1)_Q(V(1)) < d$. Clearly
$U(1)\cap V(1) = \emptyset$. Let $\hat P_1$ and $\hat P_2$ be the
projection operators for the slits $I_1$ and $I_2$.

Particle 2 is prepared with position
$X(2)_Q(W(2))$ which is classical or approximately
classical,
\begin{equation}\ [S_{X(2)}(W(2))]^2 = |(X(2)_Q(W(2)))^2 -
(X(2)^2)_Q(W(2))| < \epsilon_2,
\end{equation}  where
$\epsilon_2$ is a very small positive standard real.

Now particles 1 and 2 interact through an impulsive von Neumann
interaction Hamiltonian 
$H(1,2)_{Q}(U(1,2)) = g\cdot [X(1)\cdot\mathcal{P}(2)]_{Q}(U(1,2))$
defined on the open subset $U(1,2)$ of $\Sigma(1,2)$. Here
$\mathcal{P}(2)$ is the self adjoint operator for the momentum of
particle 2, $\hat X(1)$ is that for the position of particle 1 and $g$ is
the coupling constant that is such that $g\cdot \Delta t$ is finite for
the infinitesimal period, $\Delta t$, during which the force acts. The
solution of the Hamiltonian equations of motion for this Hamiltonian
reveals that when the interaction has ceased the position of particle 2
has changed by an amount $g\cdot \Delta t \cdot X(1)_Q(O)$, where $O$ is
an open subset of $\Sigma(1)$. 

{\bf Proposition 4}

The final position of particle 2 is approximately classical if 
$O$ is either an open subset of $U(1)$ or an open subset of $V(1)$ such
that $X(1)_Q(O)$ is almost classical. However, if $O = U(1)\cup
V(1)$ then the final position of particle 2 is not approximately classical
which means that when particle 1's quantum position covers both
slits it is not registered by the measurement particle 2.   

{\bf Proof:}

The final position of particle 2 is $X(2)_Q(W(2)) + g\cdot \Delta t
\cdot X(1)(O)$ which we will call $X(2)_f$, the corresponding
operator is $\hat X(2)_f = \hat I(1) \otimes \hat X(2) + g\cdot \Delta
t \ \hat X(1)
\otimes \hat I(2)$ and let $O(1,2)$ be the super set of $W(2)$ and $O$.

Start by assuming that $O=U(1)$ is such that $X(1)_Q(U(1))$ is
approximately classical, then
$[S_{X(2)_{f}}(O(1,2))]^2 = [S_{X(2)}(W(2))]^2 + $ 

$(g\cdot
\Delta t)^2 [S_{X(1)}(U(1))]^2 + 2g\cdot \Delta t (-X(1)_{Q}(U(1))\cdot
X(2)_{Q}(W(2)) + (\hat X(1)\otimes \hat X(2))(O(1,2))).$

The right hand side is small if both $X(2)_Q(W(2))$ and $X(1)_Q(U(1))$
are approximately classical, since the first two terms are, by definition,
and the third term is also small because approximately classical quantum
numbers are approximately homothetic,i.e., $X(2)_Q(W(2))\approx x_2
I_Q(W(2))$ and $X(1)_Q(U(1))\approx  x_1 I_Q(U(1)) $ where $x_1$ and $x_2$
are standard real numbers. 

The same argument works if $U(1)$ replaces $V(1)$. Similar arguments work
when,for example, $X(1)_Q(W(1))$ is not approximately classical but
there is an open set $O\subset W(1)$ so that $X(1)_Q(O)$ is. However the
argument does not work when
$O = U(1) \cup V(1)$ because in that case $X(1)_Q(O)$ is not
approximately classical, i.e., $[S_{X(1)}(O)]^2$ is not small. If a
measurement of position would occur, then, according to the basic
postulate 3 A, $\hat X(2)_f$ ought to become concentrated around the value
that gets registered during the process and $[S_{X(2)_{f}}(O(1,2))]^2$
would then be small. Therefore no persistent registration is likely to
occur when
$O = U(1) \cup V(1)$. Note that a similar result occurs if $O$ is centered
around say a fifty-fifty coherent superposition of states that belong to
$U(1)
$ and $V(1)$.  
\ \ $\# $

This model result implies that the quantum particle 1 may pass through
both slits simultaneously but such events are not unambiguously revealed
(or persistently registered) by the measurement particle 2 because then
the position of particle 2 is not  even approximately classical.
Nevertheless, this example shows that it is not impossible to reintroduce
in the quantum numbers approach the counterpart of classical objectivity
provided somewhere inside the chain of measurements that separates the
quantum system and the observer, a device is classical, so to say, a
quantity $\hat A$ is measured and found to have (persistent) values in the
subset
$I$. The question to know precisely at which level of the chain such a
classical measurement apparatus is present is in last resort a question
of personal interpretation. If
$I$ can be considered to be a sharp subset, relatively to subsequent
measurement devices similar to the one described in the previous section,
all of them will reveal unambiguously values contained inside $I$ and
their result will be consistent with those associated to
$\hat A$.

 What we cannot
explain at this level, and this is the deep mystery of quantum mechanics,
the essence of the yet unsolved measurement problem, is how quantum
numbers become sharp. This objectification process, or "collapse" process
ought in principle to be due simply to the interaction between the system
and the measurement apparatus but such a process, during which
superpositions are broken is not consistent with the unitarity of
Heisenberg-Schr$\ddot{o}$dinger evolution as we have shown. This point
will be briefly discussed in the conclusions.

At least, the measurment problem suggests that it is worth investigating
non-standard (non-unitary) dynmical laws. This will be done in the next
section, where we propose (speculatively) a new type of dynamics. We
shall assume that quantum particles obey the simplest generalisation of
classical dynamics that can be derived on the assumption that
quantities take  quantum real number values.

\section{Quantum Dynamics with Quantum Real Numbers.}

In the first section, we introduced sharply collimated
particles as a heuristic example which helped
to motivate the choice of the basic postulates 1 and 2.
These particles can be considered to be classical in
the sense that they behave like localised pointlike
particles with regard to passing or not through a
slit. In this section, open subsets of $\Sigma$
containing sharply collimated particles are used to
show that the unitary quantum mechanical evolution
laws give good approximations to quasi-classical dynamical laws
expressed in quantum real numbers. The set $\Sigma$
of density matrices is restricted so that the
unbounded position and momentum operators, $\hat Q_j$
and $\hat P_j$, of the Schr\"{o}dinger representation of
the canonical commutation relations give quantum real
numbers as continuous functions on $\Sigma$ \cite{adelman2}.

{\bf Basic Postulate 4 (tentative)}

Consider the example of a non-relativistic quantum particle
of positive mass $\mu$ that moves in a central force field
$F$ which is derived from a potential function $V$. We assume
that the quantum values $(Q_{j})_Q$  of the position coordinates and
$(P_{j})_Q$ of the conjugate momenta of the particle globally satisfy 
equations of motion that resemble the equations of classical mechanics.
That is, the global quantum numbers $(Q_j)_Q$ and $(P_j)_Q$ satisfy
Hamilton's equations.
Thus, $\mu\,{d(Q_j)_Q(U)\over dt}$ =  $(P_j)_Q(U)$ and
${d(P_j)_Q(U)\over dt}$ = $F_{j}(\vec Q_Q(U))$
hold for all open subsets $U \in{\Sigma}$,
where  ${d\over dt}$ denotes differentiation with
respect to time, $F_j$ represents the the $j^{th}$ component of the
force. $F_j$ is the $j^{th}$ component of the negative gradient,
${-\nabla V}$, of the scalar potential function
$V(\vec Q_Q)$ where $\vec Q_Q = ((Q_1)_Q,(Q_2)_Q,(Q_3)_Q))$.

This means that for all $\rho \in \Sigma$, 
$\mu\,{dTr(\hat Q_j \hat\rho)\over dt}$ = $ Tr(\hat P_j\hat\rho)$
and
${dTr(\hat P_j \hat\rho)\over dt}$ = $F_j(Tr(\hat Q \hat\rho))$

We will sometimes use Newton's equations which are,
in terms of the $(Q_j)_Q(\Sigma)$,
$\mu\,{d^{2}((Q_j)_Q(\Sigma))\over dt^{2}}$ = $F_j(Q_Q(\Sigma))$.
Again this means that for all $\rho \in{\Sigma}$,
$\mu\,{d^{2}(Tr(\hat Q_j \hat\rho)\over dt^{2}})$ =
$F_j(Tr(\hat Q \hat\rho))$.

{\bf An inverse to Ehrenfest's Theorem}

We will now prove a theorem that states that if
basic postulate 4 holds then the self-adjoint operators
$\hat Q_j$ and $\hat P_j$ satisfy
equations that well approximate Heisenberg's operator
equations of motion when localised to certain
open subsets of $\Sigma$. To simplify the notation we will 
assume that the particle is one dimensional.  

The theorem relates a set of operator equations,
Heisenberg's equations, to a set of quantum real number
equations, Newton's equations so we have first to explain 
what approximate equality between them means. A
straightforward way to get a quantum real number
equation from an operator
equation is to multiply each side of the operator
equation by a density operator, $\rho$, and then
take the trace of each side. The original operator
equation becomes a family of numerical equations
which can be localised in an open subset of $\Sigma$
by restricting the $\rho$'s to belong to the subset.

Recall that Heisenberg's equations for an operator
$\hat A$ are

${d\hat A\over dt}$ = $-i[\hat A , \hat H]$

where $\hat H$ is the Hamiltonian operator of the
system and the square bracket denotes the operator
commutator. For the one dimensional motion,
the Hamiltonian operator is

$\hat H$ = $1\over (2\mu)$ $ \hat P^{2}$ + $V(\hat Q)$.

To simplify the discussion we remove the explicit
dependence of the equations on the momentum operator
$\hat P$ and just use Newton's equations of motion in the form of
second order differential equations. If Newton's quantum real number
equations hold to the extent $W$, then for all
$\hat\rho$ in $W$,

$\mu\,{d^{2}(Tr(\hat Q \hat\rho)\over dt^{2}})$
= $F(Tr \hat\rho \hat Q)$.

If Heisenberg's numerical equations hold to the
extent $W$, then for all $\hat\rho$ in $W$

$\mu\,{d^{2}(Tr(\hat Q \hat\rho)\over dt^{2}})$
= $Tr \hat\rho \hat F(Q)$.

The difference between the right hand sides of
these equations shows why Ehrenfest's Theorem is 
not valid for all functions $F$. In general,

$Tr \hat\rho \hat F(Q)\ \neq  \ F(Tr \hat\rho \hat Q)$. Note that in
principle this difference is experimentally testable, which shows that
the quantum real number approach to quantum mechanics is not purely ad
hoc.

It is possible, however, that the difference
between the two sides is small at some state
$\hat\rho_{a}$ and remains small in an open
neighborhood of $\hat\rho_{a}$. Then the
equations are approximately equal on that
open set. This will be taken to mean that
Heisenberg's numerical equations give a good
approximation to Newton's equation in that
neighbourhood. We claim that for a suitable
class of functions $F$, this is true in the
vicinity of every point on the position
line of the one dimensional model.
That is, for every standard real number $r$
and standard real number  $\epsilon>0$,
we can find an open set, $W(r,\epsilon)$ in $\Sigma$, such
that, both

(a) the quantum real number  $Q_Q(W(r,\epsilon))$
is arbitrarily close to $r$, and

(b)for each $\hat\rho$ in $W(r,\epsilon)$, $Tr \hat\rho \hat F(Q)$ is
arbitrarily close to $F(Tr \hat\rho \hat Q)$.

The physical interpretation is that if an
observer's measurement apparatus is located
in the immediate vicinity of the position $r$
then the observer cannot measure any
significant difference between the accelerations
of the particle due to the two forces,
$Tr \hat\rho \hat F(Q)$ and $F(Tr \hat\rho \hat Q)$.
The unitary evolution of quantum mechanics
gives a local linear approximation to the equations
of classical mechanics expressed in quantum real
numbers.

The class of suitable functions is defined through the
concept of $\cal S$-continuity. 

{\bf Definition 3:}

A function $F$ is $\cal S$-continuous,
if it is real-valued continuous functions of a real
variable such that, for the position operator $\hat Q$,
$F(\hat Q)$ defines an operator on the Schwartz space $\bf S$ 
that is continuous in the standard countably normed topology
on $\bf S$.

The class of $\cal S$-continuous functions includes all polynomials
\cite{adelman2}.

{\bf Theorem 6:}

If the force $F$ is $\cal S$-continuous,
then given $\epsilon>0$, Heisenberg's equations of
motion approximate Newton's equations of motion to within
$\epsilon$ on each member of a collection of open sets
$W(r,\epsilon)$ of $\Sigma$,
indexed by the standard real numbers $r$ and $\epsilon$.
That is, for all $\hat\rho$ in $W(r,\epsilon)$,

$|Tr \hat\rho \hat F(Q) - F(Tr \hat\rho \hat Q)|< \epsilon$.

{\bf Proof:}

The idea behind the proof is to find states $\rho_{r}$ at
which $F(Tr \hat\rho_{r} \hat Q)$ closely approximates
$Tr \hat\rho_{r} \hat F(Q)$, then $F(Tr \hat\rho \hat Q)$
will be close to $Tr \hat\rho \hat F(Q)$ for all $\hat\rho$
that are such that both
$F(Tr \hat\rho \hat Q)$ is close to
$F(Tr \hat\rho_{r} \hat Q)$ and
$Tr \hat\rho \hat F(Q)$ is close to
$Tr \hat\rho_{r} \hat F(Q)$.
To achieve this we must first construct the open sets
$W(r,\epsilon)$.

{\bf Definition 4:}

Given $F$,$r$ and $\epsilon$, $W(r,\epsilon)$ = $N(\hat\rho_{r},\hat
Q,\delta)\cap N(\hat\rho_{r},\hat F(Q),{\epsilon\over 3})$,
where, $\delta$ is given by
$|F(r)-F(x)|< {\epsilon\over 6}$ if $|r-x|<\delta$
($\delta$ depends upon both $\epsilon$ and $r$) and
$\rho_{r}$ satisfies both
$|Tr\hat\rho_{r} \hat F(Q) - F(r)|< {\epsilon\over 6}$
and
$|Tr\hat\rho_{r} \hat Q - r|< {\delta \over 2}$.

That such density operators $\rho_{r}$ exist follows from
Weyl's criterion \cite{reed}.

{\bf Lemma 1:}

If $Q$ is a self-adjoint operator which has
absolutely continuous spectrum,
then for any real number $r$
in its spectrum we can construct a sequence of pure states
$\hat\rho_{n}$ such that, for the given $S$-continuous
function $F$,
$Tr\hat\rho_{n} \hat F(Q)$ approaches $F(r)$
and  $Tr\hat\rho_{n} \hat Q$ approaches $r$,
as n approaches infinity.

{\bf Proof:}

From Weyl's criterion \cite{reed} it follows that,
for any number $r$ in the spectrum of $\hat Q$, there exists a
sequence of unit vectors $\{u_{n}\}$, in the domain of $\hat Q$,
such that if $\hat\rho_{n}$ is the projection onto the
one dimensional subspace spanned by the vector $u_{n}$
then $Tr\hat\rho_{n} \hat Q$ approaches $r$ as $n \to \infty$.

The vectors $\{u_{n}\}$ can be chosen to be in $\bf S$.
Furthermore we can find a sequence of vectors
$\{u_{n}\} \in \bf S$ such that for $n$ large enough the support
of $u_{n}$ lies in a narrow interval centred on $r$.
Then, the corresponding one dimensional projection operators
$(\hat\rho_{n})$, are such that the sequence of standard
real numbers $Tr\hat\rho_{n} \hat F(Q)$ approaches $F(r)$
by S-continuity, and the sequence of standard real numbers
$Tr\hat\rho_{n} \hat Q$ approaches $r$ by the spectral
theorem for $Q$.
$\#$

From Lemma 1, once we are given a real number $r$ in the
spectrum of $Q$, the $\cal S$-continuous function $F$ and a
real number $\epsilon>0$, we can find an integer $N$ such
that, for all $j>N$, both
$|Tr\hat\rho_{j} \hat F(Q) - F(r)|< {\epsilon\over 6}$
and
$|Tr\hat\rho_{j} \hat Q - r|< {\delta\over 2}$
where $\delta$ is given in Definition 4.

We choose $\hat\rho_{r}$ =$\hat\rho_{j}$, for some $j>N$, and
deduce that

$|Tr\hat\rho_{r} \hat F(Q) - F(Tr\hat\rho_{r} \hat Q)|
< {\epsilon\over 3}$

because

$|Tr\hat\rho_{r} \hat F(Q) - F(Tr\hat\rho_{r} \hat Q)|$

 $\leq |Tr\hat\rho_{r} \hat F(Q) - F(r)|$
+ $|F(r) - F(Tr\hat\rho_{r} \hat Q)|$.

With this choice of $\rho_{r}$, the construction of
the open set $W(r,\epsilon)$ is completed.

{\bf Proof of Theorem 6, (continued):}

For all $\rho \in W(r,\epsilon)$ we have

$|Tr\hat\rho \hat F(Q) - F(Tr\hat\rho \hat Q)|$

$\leq |Tr\hat\rho \hat F(Q) - Tr\hat\rho_{r} \hat F(Q)|$
+ $|Tr\hat\rho_{r} \hat F(Q) - F(Tr\hat\rho_{r} \hat Q)|$
+ $|F(Tr\hat\rho_{r} \hat Q)-F(Tr\hat\rho \hat Q)|$.

If $\rho$ is in $N(\hat\rho_{r},\hat F(Q),{\epsilon\over 3})$
the first summand is $< {\epsilon\over 3}$, as is the second
by choice of $\rho_{r}$. The final summand is also
$<{\epsilon \over 3}$ because

$|F(Tr\hat\rho_{r} \hat Q)-F(Tr\hat\rho \hat Q)| \leq
|F(Tr\hat\rho_{r} \hat Q)- F(r)|+|F(r)- F(Tr\hat\rho \hat
Q)|$.

Here the first summand is $< {\epsilon \over 6}$ by the
definition of $\hat\rho_{r}$ in Definition 4.
Furthermore,the second summand is $< {\epsilon \over 6}$
because the function
$F$ is continuous at $r$ and with $x=Tr\hat\rho \hat Q$,
$|x-r|\leq |x-Tr\hat\rho_{r}\hat Q| +|Tr\hat\rho_{r}\hat Q
-r| < {\delta \over 2} + {\delta \over 2} = \delta$,
because $\hat \rho \in W(r,\epsilon)$, and
because  of the choice of $\hat \rho_r$ in Definition 4.

Therefore, for any $\rho$ in $W(r,\epsilon)$,
$|Tr\hat\rho \hat F(Q) - F(Tr\hat\rho \hat Q)|$
$< \epsilon$. $\#$

The question remains whether we can construct sufficiently
many of these open sets. In general, for a given $S$-smooth
function $F$, the family of open sets
$\bigl\{W(r,\epsilon)\bigr\}$, does not form an open cover of
the state space $\Sigma$. The physically important exception is
the linear force law,eg simple harmonic motion, when equality
holds for all $\rho$ in $\Sigma$.

However, for every permissible $F$, the family of open
sets $\bigl\{W(r,\epsilon)\bigr\}$ covers the classical
coordinate space of the physical system in the sense
that associated with each $W(r,\epsilon)$ there is an
open interval $(r-\delta,r+\delta)$, with $\delta $ defined in Definition
4,such that the collection of these intervals covers the standard real
line which is the classical coordinate space of this model.

If we had used three dimensions for the classical configuration
space of the particle, the analog of Theorem
6 would permit us to deduce that there is a
family of open sets $\bigl\{W(\vec{x},\epsilon)\bigr\}$,
with $\vec{x}\in \bf R^{3}$,on which Heisenberg's
numerical equations give a good approximation to
Newton's equation. Furthermore associated to each
$W(\vec{x},\epsilon)$ is an open ball
$B(\vec{x},\delta)$ in $\bf R^{3}$, the collection of which cover
$\bf R^{3}$. Again an observer measuring a particle
with apparatus set up in one of these open balls
could not determine locally whether the evolution
of the particle was governed by Heisenberg's equations
of motion averaged over a $\rho$ from
$W(\vec{x},\epsilon)$ or by Newtons equations of motion
for the quantum numbers $\vec{Q}\mid_{W}$ restricted to
$W(\vec{x},\epsilon)$.

It is interesting to see how these results correlate with
the ideas of collimation of a particle. Take the open interval
$(z_1,z_2) = (r-\delta, r + \delta)$ to be the slit
through which the particle is  $\epsilon$ sharp
collimated. Let $U$ be such that for all $\rho\in{U}$,
the particle is $\epsilon$-sharp collimated, if
$\delta < m\cdot \epsilon$, where $m$ = $2\cdot min(|z_1|, |z_2|)$ (see
Theorem 4)
and
$\hat Q$ is taken to be $\hat Z$ then $W(r,\epsilon)$ is contained in $U$,
to an extent that  depends on the force $F$.

\section{\bf Conclusions and remarks.}
\subsection{\bf Some remarks.}

{\bf a) The importance of continuity.}

It is worth noting that continuity
is the basic property that allows a physical theory to remain valid under
slight changes (up to an $\epsilon$) and even to be persistent in the
presence of profound reformulations.
In our approach, continuity was present at all levels: the deduction of the
form of the unitary evolution laws and of the quantum probability rule
are based on a requirement of continuity between the classical and the
quantum regimes (the law of large numbers itself presupposes
some kind of continuity). The basic postulate 3 reflects at the quantum
level the classical properties of continuity in time of the physical
magnitudes (persistence). The central role played by continuity
is too often neglected or ignored in quantum mechanics. Our formulation in
terms of quantum numbers helps to restore the centrality of the role
played by continuity in quantum mechanics.

{\bf b) About quantum paradoxes.}

Let us now quickly look at three celebrated paradoxes, the EPR, the
Schr$\ddot{o}$dinger cat and the quantum Zeno paradoxes using the language of
quantum real numbers.

Provided we think in terms of quantum real numbers,
the values taken by physical properties can always be expressed as quantum
numbers and only in extreme circumstances, such as an $\epsilon$ sharp
collimation, are well-approximated by standard real numbers. So we must
abandon that part of our classical intuition according to which the values
of quantities preexist as standard real numbers before the measurement.
They only pre-exist as quantum real numbers. In our model different
classical standard real number values of position can be determined by
measurements
on a single particle with a single quantum real number value for its position.
Therefore the usual concept of localisation which refers to classical
standard real
number values of position needs to be reviewed in the light of the particle
having quantum real values for its position. The particle may be localised
in terms of the quantum real number values of its position but not localised
in terms of the classical standard real number values of its position. We
plan to
examine the Einstein- Podolsky- Rosen paradox in detail in a future work.

 A cat composed of atoms and molecules which may be localised in terms of
quantum real number values but not localised in terms of classical standard
real
number values could as well be both living and dead to an observer if
the difference between being alive and being dead is just a question of
molecular configurations.  In the standard theories of quantum mechanics
there remains the basic problem of the quantum theory of measurement:
to precisely determine the border-line that separates  a measurement
regime from a regime of unitary evolution. What is it that actualises
potentialities? Our model provides a different way of posing the question.
Do there exist Newtonian forces that when expressed in quantum real
numbers allow a quantity whose quantum real number values are not
$\epsilon$ sharp collimated to evolve so that its quantum real number
values do become $\epsilon$ sharp collimated? It seems reasonable that
such forces exist, but we do not yet have an answer to this question.
Note that an equation of the kind $\mu\,{d(Tr(\hat Q \hat\rho)\over
dt})$ = $\lambda \cdot(Tr^2 \hat\rho \hat Q-Tr \hat\rho \hat Q^2)$
(with $\lambda$ taken to be a positive real) makes it possible to describe
the measurement (sharpening) of the observable
$\hat Q$. However, such an evolution is nor an Hamiltonian evolution
(because it introduces an arrow of time) neither a Newtonian one because
it does not contain any acceleration term. It is certainly not a
Schr$\ddot{o}$dingerlike, unitary, evolution because it is not linear in
$\hat\rho$. Nevertheless it is expressed solely in terms of quantum real
numbers.

The quantum Zeno paradox is based on the assumptions that in a measurement
the wave function collapses and that the collapse process is
instantaneous. Firstly, it is worth noting that if a chain
of measuring devices is present between the observed quantum system and
the human observer, as is always the case, it is clear that the
measurement process is not instantaneous, a point that was made
clear through our analysis of the impulsive von Neumann interaction.
 Beside, as we have  suggested following our analysis of the
proposition 4, there ought to exist dynamical forces that link the
quantum real number values of physical quantities to classical values of
quantities associated with the measurement apparatus and thereby cause
the quantum real number values to become $\epsilon$ sharp. Such forces
could be used to model measurement processes to give a quasi-dynamical
description of the "collapse" in which there will be no Zeno paradox.
Moreover it can be shown that, in virtue of the law of large numbers,
when the number of particles of the system under observation increases,
the effect of the collapse process decreases proportionnally because the
Hilbertian distance between the initial state and the collapsed state
decreases when
$N$ increases. Then, provided the measurement time $\tau$ is very small
but not negligibly small, the change imposed during a time $T$  by a
series of $T\over \tau$ successive measurements will become negligible in
the limit of large numbers. This property is an extension of the results
derived in the section 2. We plan to examine it in detail in a future
work.

This section has not provided hard solutions to the paradoxes of quantum
mechanics but it does outline some projects of using quantum real numbers
to study and perhaps resolve them.

\subsection{\bf Conclusions.}
In order to clarify the correspondences between conventional
quantum mechanics and the quantum numbers approach,we will 
compare the standard axioms of quantum mechanics as they are enumerated
in the text-book of Cohen-tannoudji {\em et al.}
\cite{cohen-tannoudji} and our basic postulates:

Standard Axioms 1, 2 and 3: 1; states are represented by rays of the
Hilbert space or convex combinations of them (density
matrices), 2; measurable quantities are represented by
self-adjoint operators (observables) and 3; measurable values are
eigenvalues of these
operators (in other words, observed
physical quantities belong to the spectrum of the observable
under measurement).

Basic postulate 1; Physical quantities always have numerical values as
quantum real numbers of the form
$M_Q(U)\ =\ Tr(\hat \rho.\hat M)_{\hat
\rho\
\in\
U}$ where $U$ is an open subset of
the set of density matrices $\Sigma $,
and $\hat M$ is a self-adjoint, continuous linear operator.Any system
always has an open set of states associated with it. The measured values
of a physical quantity always belong to the spectrum of the corresponding
self-adjoint operator.

The standard axiom 4 in ref.\cite{cohen-tannoudji} is the
Born rule.

The Born rule is a consequence of 
the basic postulate 1 and basic postulate 2, the ergodicity assumption. 

The standard axiom 5\cite{cohen-tannoudji} is the collapse hypothesis. 

The collapse hypothesis, in the form of the Luders-von Neumann
transformation rule (so to say in its weakest form), is a consequence of
the Born rule and of our postulates 3 A and 3 B, which characterize the
persistence in time of observed outcomes.

The standard axiom 6\cite{cohen-tannoudji} assumes that the time
evolution is given by the Schr$\ddot{o}$dinger equation or equivalently
the Heisenberg equations for the observables.

In a dynamical model of the measurement of the position
of a particle we showed how the quantum real number value of the
position is forced to be an almost classical real number if we impose
that it gets registered during a unitary interaction with the classical
measurement apparatus, which suggests that new, non-unitray
dynamics ought to be studied in the framework of the quantum real number
interpretation of quantum mechanics.

Basic postulate 4 states that the position and momentum of a particle
when expressed in quantum real numbers satisfy Hamiltonian/Newtonian
equations of motion. Theorem 6 shows that for a certain class of
forces, there are open sets of state space on which Heisenberg's
equations of motion give close approximations to Newton's equation of
motion in quantum real numbers. Furthermore while this class of open
sets doesn't cover state space, it does cover the classical position
space of the particle. Thus at every point in position space we
cannot distinguish locally between the two types of dynamical
motion. 

We hope that the quantum real number interpretation of quantum mechanics
will open the way for a deeper understanding.

\section*{Acknowledgments}T.D. is a
Postdoctoral Fellow of the Fonds voor Wetenschappelijke Onderzoek,
Vlaanderen.


\begin{thebibliography}{99}
\bibitem{adelman1} M.Adelman and J. V. Corbett: "A Sheaf Model
for Intuitionistic Quantum Mechanics", Applied Categorical Structures {\bf
3} 79-104 (1995)\smallskip
\bibitem{adelman2} M.Adelman and J.V. Corbett: "Quantum Mechanics
as an Intuitionistic form of Classical Mechanics" to appear in
Proceedings of the Centre Mathematics and its Applications, ANU,
Canberra (2001).\smallskip
\bibitem{ballentine} L. E. Ballentine: ``Limitations of the projection
postulate'', Founds.
Phys. {\bf 20} n$^{\circ}$ 3 (1990) 1329.\smallskip
\bibitem{cohen-tannoudji}C. Cohen-Tannoudji, B. Diu and F. Laloe, {\em
Mecanique quantique},
(Hermann, Paris, 1977).\smallskip
\bibitem{birkhoff1} G. Birkhoff and J. von Neumann: ``The logic of
quantum mechanics'', Annals of Mathematics 37 (1936) 823.\smallskip
\bibitem{busch} P. Busch and P. Lahti: ``Individual aspects of quantum
measurements'' J. Phys. A:
Math.Gen. {\bf 29} (1996)\smallskip
\bibitem{connes} A. Connes, {\em Geometrie Non-Commutative},
(InterEditions,Paris,
1990).\smallskip

\bibitem{dewndney} C. Dewndney, P.R. Holland, A. Kyprianidis and J.P. Vigier:
``Spin and non-locality in quantum mechanics'', Nature {\bf 336}
n$^\circ$6199 (1988)
536 \smallskip
\bibitem{goldstone} E. Farhi, J. Goldstone and S. Gutmann: ``How
probability arises in quantum
mechanics'' Ann. Phys. (NY){\bf 192} (1989).\smallskip
\bibitem{heisen} W. Heisenberg, Zs.Phys.{\bf 33} (1925) pp879-893,
translated in
B. L. van der Waerden, {\em Sources of Quantum Mechanics}, pp261-276
(Dover, 1968).\smallskip
\bibitem{heyting} A. Heyting: ``Intuitionism, an introduction'' eds. North
Holland, Amsterdam.\smallskip
(1971).

\bibitem{gutko} D. Gutkoski and G. Masotto ``An inequality stronger than
Bell's inequality'', {\it Nuov. Cim.}, {\bf 22 B},  n$^\circ$1,
(1974).\smallskip

\bibitem{lindsay} R. B. Lindsay  and H. Margenau, {\em Foundations of Physics},
p397 (Dover, 1957). \smallskip
\bibitem{maclane}S. MacLane and I. Moerdijk, {\em Sheaves in Geometry
and Logic} (Springer--Verlag, New York, 1994).\smallskip
\bibitem{reed}W. Reed and B. Simon, {\em Methods of Mathematical
Physics I: Functional Analysis} (Academic Press, New York,
1972).\smallskip
\bibitem{squires} J. Squires: ``On an alleged ``proof'' of the quantum
probability law'' Phys. Lett.
A {\bf 145}n$^\circ$23 (1990).\smallskip
\bibitem{stout}L. N. Stout, Cahiers Top. et Geom. Diff. {\bf XVII},
295 (1976); C. Mulvey, ``Intuitionistic Algebra and Representation
of Rings'' in Memoirs of the AMS {\bf 148} (1974).\smallskip
\bibitem{gisin} W. Tittel, J. Brendel, H. Zbinden and N. Gisin:
``Long-distance Bell-type
tests using energy-time entangled photons'', Phys. Rev. A {\bf 59} (1999)
4150.\smallskip
\bibitem{vonneuneu2} J. V. von Neumann: ``Mathematische grundlagen
der quanten-mechanik'',  Springer-Verlag, Berlin (1932).\smallskip

\end{thebibliography}

\end{document}